\documentstyle[psfig]{mn}

\newif\ifAMStwofonts

\def\lsim{\raise0.3ex\hbox{$<$}\kern-0.75em{\lower0.65ex\hbox{$\sim$}}}
\def\gsim{\raise0.3ex\hbox{$>$}\kern-0.75em{\lower0.65ex\hbox{$\sim$}}}

\ifoldfss
  \ifCUPmtlplainloaded \else
    \NewTextAlphabet{textbfit} {cmbxti10} {}
    \NewTextAlphabet{textbfss} {cmssbx10} {}
    \NewMathAlphabet{mathbfit} {cmbxti10} {} 
    \NewMathAlphabet{mathbfss} {cmssbx10} {} 
  \fi
  \ifAMStwofonts
    \ifCUPmtlplainloaded \else
      \NewSymbolFont{upmath} {eurm10}
      \NewSymbolFont{AMSa} {msam10}
      \NewMathSymbol{\upi}     {0}{upmath}{19}
      \NewMathSymbol{\umu}     {0}{upmath}{16}
      \NewMathSymbol{\upartial}{0}{upmath}{40}
      \NewMathSymbol{\leqslant}{3}{AMSa}{36}
      \NewMathSymbol{\geqslant}{3}{AMSa}{3E}

       \let\le=\leqslant
       \let\ge=\geqslant
    \fi
  \fi
\fi 

\ifnfssone
  \newmathalphabet{\mathit}
  \addtoversion{normal}{\mathit}{cmr}{m}{it}
  \addtoversion{bold}{\mathit}{cmr}{bx}{it}
  \newmathalphabet{\mathbfit} 
  \addtoversion{normal}{\mathbfit}{cmr}{bx}{it}
  \addtoversion{bold}{\mathbfit}{cmr}{bx}{it}
  \newmathalphabet{\mathbfss} 
  \addtoversion{normal}{\mathbfss}{cmss}{bx}{n}
  \addtoversion{bold}{\mathbfss}{cmss}{bx}{n}
  \ifAMStwofonts
    \ifCUPmtlplainloaded \else
      \UseAMStwoboldmath
      \makeatletter
      \new@mathgroup\upmath@group
      \define@mathgroup\mv@normal\upmath@group{eur}{m}{n}
      \define@mathgroup\mv@bold\upmath@group{eur}{b}{n}
      \edef\UPM{\hexnumber\upmath@group}
      \new@mathgroup\amsa@group
      \define@mathgroup\mv@normal\amsa@group{msa}{m}{n}
      \define@mathgroup\mv@bold\amsa@group{msa}{m}{n}
      \edef\AMSa{\hexnumber\amsa@group}
      \makeatother
      \mathchardef\upi="0\UPM19
      \mathchardef\umu="0\UPM16
      \mathchardef\upartial="0\UPM40
      \mathchardef\leqslant="3\AMSa36
      \mathchardef\geqslant="3\AMSa3E

       \let\le=\leqslant
       \let\ge=\geqslant
    \fi
  \fi
\fi 

\ifnfsstwo
  \DeclareMathAlphabet{\mathbfit}{OT1}{cmr}{bx}{it}
  \SetMathAlphabet\mathbfit{bold}{OT1}{cmr}{bx}{it}
  \DeclareMathAlphabet{\mathbfss}{OT1}{cmss}{bx}{n}
  \SetMathAlphabet\mathbfss{bold}{OT1}{cmss}{bx}{n}
  \ifAMStwofonts
    \ifCUPmtlplainloaded \else
      \DeclareSymbolFont{UPM}{U}{eur}{m}{n}
      \SetSymbolFont{UPM}{bold}{U}{eur}{b}{n}
      \DeclareSymbolFont{AMSa}{U}{msa}{m}{n}
      \DeclareMathSymbol{\upi}{0}{UPM}{"19}
      \DeclareMathSymbol{\umu}{0}{UPM}{"16}
      \DeclareMathSymbol{\upartial}{0}{UPM}{"40}
      \DeclareMathSymbol{\leqslant}{3}{AMSa}{"36}
      \DeclareMathSymbol{\geqslant}{3}{AMSa}{"3E}

       \let\le=\leqslant
       \let\ge=\geqslant
    \fi
  \fi
\fi 

\ifCUPmtlplainloaded \else
  \ifAMStwofonts \else 
    \def\upi{\pi}
    \def\umu{\mu}
    \def\upartial{\partial}
  \fi
\fi

\title[Intermediate-redshift clusters]{The galaxy population of intermediate-redshift 
clusters}
\author[Dahl\'{e}n et al. ]{Tomas Dahl\'{e}n\thanks{E-mail: tomas@astro.su.se. Present address: Space Telescope Science Institute, 3700 San Martin Dr., Baltimore, MD 21218} 
, Claes Fransson, G\"{o}ran \"{O}stlin \& Magnus N\"{a}slund \\ Stockholm Observatory, Stockholm University, SE--106 91 Stockholm, Sweden}
\date{Accepted 2004 January 16. Received 2004 January 15; in original from 2002 June 19
}
\pagerange{\pageref{firstpage}--\pageref{lastpage}}
\pubyear{2004}

\begin{document}

\maketitle

\label{firstpage}
\begin{abstract}
Using photometric redshifts we determine the galaxy population of the clusters of galaxies 
Cl0016+16 at $z~=~0.55$, Cl1600+41 at $z~=~0.54$, 
Cl1601+42 at $z~=~0.54$ and 
MS1008--1224 at $z~=~0.31$. Comparing the clusters, we find no evidence for a universal 
shape of the total luminosity function (LF) at these redshifts. When dividing the LFs
into spectral types, we find that the LF of the early-type 
galaxies alone can be described by a Gaussian, while the LF of the 
late-type galaxies is well fitted by a Schechter function, suggesting that
 the separate LFs for
different populations may be universal. The difference in the 
total LFs can mainly be attributed to the varying relative normalisation of these populations, 
implying that clusters with an abundant population of late-type 
galaxies also have steeper faint-end slopes. In MS1008--1224 we detect a
faint blue population that dominates over a population with colours consistent with
dwarf ellipticals, opposite to clusters at lower redshift.

Compared to low-redshift clusters, we find that a general fading of the late-type 
population by $\sim~2$ mag and the early-type population by $\sim$ 1 mag describes the evolution from $z~=~0.55$ to $z~=~0$ well. 

As a consequence of the different early-type and late-type LFs and their 
dependence on cluster radius, the fraction of blue cluster galaxies, as 
measured by the Butcher--Oemler effect, 
differs between the clusters and depends on limiting magnitude and radius. 

We find a correlation between the dwarf-to-giant ratio and the surface 
density, indicating that the high density environment in the cluster cores is 
hostile to dwarf galaxies.

\end{abstract}
\section{Introduction}
Ever since the results of Butcher \& Oemler (1984, hereafter BO84), showing that the 
fraction of blue galaxies in clusters increases rapidly between $z~=~0$ and $z~\sim~0.5$, 
it has been clear that there is strong evolution in cluster galaxy populations with redshift. 
These results have been confirmed by subsequent surveys (e.g. Rakos \& Schombert 
1995; Ellingson et al. 2001; Kodama \& Bower 2001; Margoniner et al. 2001). A general 
picture explaining the over-all features of cluster formation and evolution is given 
within the cold dark matter (CDM) 
hierarchical clustering scenario (e.g. Bower 1991; Kauffmann \& White 1993; 
Kauffmann 1995; Baugh, Cole \& Frenk 1996). A large number of observations and N-body 
simulations now support this scenario. The blueing with redshift of the cluster galaxies 
is explained by a higher accretion rate of star forming field galaxies at higher 
redshift (Bower 1991), possibly combined with a general increase in star formation in field 
galaxies at higher $z$ (Diaferio et al. 2001). A consequence of the CDM model 
is that high-$z$ 
clusters assemble during a much shorter time interval than comparably rich clusters today 
(Kauffmann 1995). This general scenario is also supported by the observed increase of mergers in 
high-redshift clusters (van Dokkum et al. 1999). Furthermore, N-body 
simulations (e.g., Dubinski 1998) show that the brightest cluster galaxies 
(BCG) naturally form via mergers in a hierarchical scenario. 

The difference in radial distribution between early-type and late-type galaxies 
in clusters can be explained by the hierarchical model if infalling field galaxies, which are 
predominantly blue star-forming galaxies, have their star formation truncated as 
they fall deeper into the cluster potential (Balogh, Navarro \& Morris 2000; 
Diaferio et al. 2001). Mergers and interactions in the high-density cluster 
environment also affect the star formation, transforming the spectral type
of the galaxies.

One of the most important probes of cluster evolution is the luminosity function (hereafter LF), 
describing the number of galaxies per magnitude bin in a cluster. 
Binggeli, Sandage 
\& Tammann (1988) showed that the total LF for the Virgo cluster
rises as a Gaussian at 
bright magnitudes and then flattens to a "plateau" (or even a slight decrease), before 
it turns steep at faint magnitudes. LFs with similar shapes are found in other nearby 
clusters, e.g. Coma (Trentham 1998a). The existence of a numerous 
population of faint dwarf galaxies in clusters, more common than in the 
field, is verified by a number of observations (e.g., Smith, Driver \& Phillipps 
1997; Wilson et al. 1997; Phillipps et al. 1998; Trentham 1998a; Yagi et al. 2002). There are, 
however, also clusters that do not show this steep increase at faint magnitudes (Trentham 
1998b, 1998c). Most likely, this is a consequence of different evolution 
depending on e.g., richness and epoch of formation.   

Conselice, Gallagher \& Wyse (2001) discuss different scenarios 
that could lead to the dwarf population in local clusters. By comparing models with 
the observed kinematic and spatial properties of a number of Virgo cluster galaxies, 
they suggest that a transformation of spirals into dwarf ellipticals (dEs) by "galaxy harassment" (Moore, 
Lake \& Katz 1998) is occurring. Some dEs may, however, have formed outside the cluster 
in galaxy groups, which were later accreted. These dEs may therefore 
be as old as the cluster ellipticals. The fading of these cluster galaxies into todays dwarf 
population is discussed by Wilson et al. (1997) and Smail et al. (1998). 

Besides measuring 
the total LF, it is of high interest to investigate the LFs of different galaxy 
populations and their evolution with redshift. While previous studies of type-dependent 
LFs have focused on {\it morphological types}, e.g., Binggeli et al. 
(1988) and Andreon (1998), here we divide galaxies into {\it spectral types}
characterised by their colours, as described in Section 3.3. 
In this paper, when we refer to galaxies as early-types and late-types we
mean their spectral type, while morphological types are denoted by their type in the 
Hubble sequence, e.g., ellipticals and spirals.

A consequence of the hierarchical scenario is that clusters observed at high redshift 
($z~\gsim~0.5$), where the infall of field galaxies is assumed to peak (Bower, Kodama \& 
Terlevich 1998), should have a higher fraction of late-type galaxies compared to local clusters. 
This should lead to a differential evolution between the late-type and early-type 
cluster LFs with redshift, with the late-type LF shifted towards brighter 
magnitudes at increasing redshift, as compared to the early-types.

Because it is observationally costly, if at all possible, to obtain redshifts for 
the numerous faint population, most observational efforts to study galaxy populations 
in high-$z$ clusters have concentrated on the brighter galaxies ($M_B~\lsim~-19$). 
Recently, however, photometric redshifts have been demonstrated to be a powerful tool for 
studying especially the faint population of cluster galaxies. Even though the 
accuracy of photometric redshifts can not be compared to spectroscopic 
redshifts, they can provide a reliable determination of clusters membership, 
and therefore significantly reduce the necessary amount of field 
galaxy subtraction. In a previous paper 
(Dahl\'{e}n, Fransson \& N\"{a}slund 2002, hereafter DFN02), we demonstrated the use 
of photometric redshifts in a study of the population of the intermediate rich cluster 
Cl1601+42 at $z~=0.54$. 
The photometric redshift selection minimizes the amount of background subtraction 
needed. Internal properties of the cluster, such as radial distribution and luminosity 
function, both for the total cluster population, as well as for different populations 
separately, can therefore be determined. Here we extend our previous study of Cl1601+42 
to include two additional rich clusters, Cl0016+16 at $z~=~0.55$ and MS1008--1224 at $z~=~0.31$. 
We also include observations of a fourth poor cluster, Cl1600+41 at $z~=~0.54$. 

Throughout this paper we assume a Hubble constant $H_0~=50$ km s$^{-1}$ Mpc$^{-1}$, 
and a cosmology with $\Omega_M~=~0.3$ and $\Omega_{\Lambda}~=~0.7$, unless anything else
is assumed. Magnitudes 
are given in the Vega based system.

\section{The Data}
\subsection{Observations}
The positions of the four clusters and a blank field, used for background subtraction, are 
listed in Table \ref{Table1}.
\begin{table*}
\begin{minipage}{140mm}
\caption{Positions of the observed clusters and back ground field given in J2000 
coordinates.}

\begin{tabular}{lcc}
\bf Object &\bf RA & \bf Dec \\ 
& &\\
Cl0016+16   & 00$^h$18$^m$33$^s$.3  & $16^{\circ}$~26\arcmin~36\arcsec \\ 
Cl1600+41   & 16$^h$02$^m$06$^s$.1  & $41^{\circ}$~01\arcmin~23\arcsec \\ 
Cl1601+42   & 16$^h$03$^m$09$^s$.8  & $42^{\circ}$~45\arcmin~18\arcsec \\ 
MS1008-1224 & 10$^h$10$^m$34$^s$.1 & $-12^{\circ}$~39\arcmin~48\arcsec \\ 
Blank field & 16$^h$08$^m$54$^s$.0  & $41^{\circ}$~34\arcmin~00\arcsec \\ 
\end{tabular}
\label{Table1}
\end{minipage}
\end{table*}
All observations, except those of MS1008--1224, were carried out with the 2.56-m 
Nordic Optical Telescope (NOT) and the Andalucia Faint Object Spectrograph and Camera 
(ALFOSC) during six observing runs between 1997 and 2001. The clusters and the background field 
were observed in four filters, $B$, $V$, $R$ and $I$, and additional observations in the 
$U$ filter were obtained for Cl1601+42. Observations were performed under photometric 
conditions. The seeing in the images varies between 0\arcsec.70 and 1\arcsec.15. A 
complete log of the observations is given in Table \ref{Table2}.

The observations of MS1008--1224 were carried out by the Science Verification 
Team at ESO using FORS at the VLT. The cluster was observed in £$B,V,R,I$ FORS 
Bessel filters with exposure times $\sim~1-1.5$h. The images were reduced using 
the {\sc iraf} package. The seeing in the final coadded frames is
(0\arcsec.72, 0\arcsec.65, 0\arcsec.64, 0\arcsec.55) in the ($B,V,R,I$) filter. 
Further details can be found at the ESO web 
site\footnote{URL http://www.hq.eso.org/science/ut1sv}. 
\begin{table*}
\begin{minipage}{140mm}
\caption{Log of observations.}
\begin{tabular}{lclrccc}
\bf Object &\bf Filter & \bf Obs. date & \bf Exp. time & {\bf \# of exp.} & \bf Seeing 
& \bf 1$\sigma$ (mag arcsec$^{-2}$) \\ 
& & & & & \\
Cl0016+16 & $B$ & Aug 00 & 18000s & 20 & 0\arcsec.92 & 27.7\\ 
 & $V$ & Aug 00 & 17100s & 19 & 1\arcsec.03 & 27.3\\ 
 & $R$ & Aug 00 & 7200s & 8 & 0\arcsec.73 & 26.6\\
 & $I$ & Aug 00 & 16200s & 36 & 0\arcsec.74 & 26.0\\ 
 
Cl1600+41 & $B$ & Jun 01 & 18000s & 20 & 0\arcsec.89 & 27.8\\ 
 & $V$ & Jun 01 & 12600s & 14 & 0\arcsec.86 & 27.2\\ 
 & $R$ & Jun 01 & 8100s & 9 & 0\arcsec.89 & 26.7\\
 & $I$ & Jun 01 & 18000s & 30 & 0\arcsec.83 & 26.2\\ 
 
Cl1601+42 & $U$ & Apr/Jun 99 & 25200s & 14 & 0\arcsec.89 & 26.7\\ 
 & $B$ & Jun 98 & 12600s & 14 & 0\arcsec.84 & 27.5\\ 
 & $V$ & Apr 99 & 16200s & 18 & 1\arcsec.02 & 27.3\\ 
 & $R$ & Jun 97 & 7200s & 8 & 0\arcsec.75 & 26.8\\
 & $I$ & Jun 98/Jun 99 & 14400s & 24 & 0\arcsec.76 & 26.2\\ 
 
Blank field & $B$ & Aug 00 & 16200s & 18 & 1\arcsec.15 & 27.6\\
 & $V$ & Jun 99 & 16200s & 18 & 1\arcsec.02 & 27.0\\ 
 & $R$ & Jun 97 & 7200s & 8 & 0\arcsec.70 & 26.8\\
 & $I$ & Aug 00 & 16200s & 36 & 0\arcsec.73 & 26.8\\ 
\end{tabular}
\label{Table2}
\end{minipage}
\end{table*}
Information on redshifts, galactic extinction, observed area, X-ray luminosities and 
velocity dispersion of the clusters is given in Table \ref{Table3}.
\begin{table*}
\begin{minipage}{140mm}
\caption{Properties of the cluster sample.}
\begin{tabular}{lcccccc}
\bf Cluster &\bf Redshift &\bf $A_V^a$ &\bf Observed area &\bf Diameter$^b$ &{\bf $L_x$} 
(0.3-3.5 keV) & \bf $\sigma$ \\ 
 & & & sq. arcmin & Mpc & $10^{44}$ erg s$^{-1}$ & km s$^{-1}$ \\
\hline
Cl0016+16 & 0.546$^c$ & 0.19 & 32.5 & 3.1 & 34.7$^d$ & 1703$^d$ \\
Cl1600+41 & 0.540$^i$ & 0.04 & 28.9 & 2.9 & $<$2.0$^{i,j}$ & - \\
Cl1601+42 & 0.539$^e$ & 0.03 & 30.4 & 3.0 & 2.1$^d$ & 1166$^d$ \\
MS1008-1224 & 0.306$^f$ & 0.23 & 32.4 & 2.2 & 5.8$^g$ & 1054$^h$ \\
\hline
\multicolumn{6}{l}{Notes:}\\
\multicolumn{6}{l}{$a$) Schlegel et al. 1998}\\
\multicolumn{6}{l}{$b$) Cluster diameter covered by observations.}\\
\multicolumn{6}{l}{$c$) Dressler \& Gunn 1992}\\
\multicolumn{6}{l}{$d$) Smail et al. 1997.}\\
\multicolumn{6}{l}{$e$) Oke, Gunn \& Hoessel 1996.}\\
\multicolumn{6}{l}{$f$) Lewis et al. 1999.}\\
\multicolumn{6}{l}{$g$) Gioia \& Luppino 1994.}\\
\multicolumn{6}{l}{$h$) Carlberg et al. 1996.}\\
\multicolumn{6}{l}{$i$) Henry et al. 1982.}\\
\multicolumn{6}{l}{$j$) Converted from energy band 0.5-4.5 keV assuming electron 
temperature 4 keV}\\

\hline
\end{tabular}
\label{Table3}
\end{minipage}
\end{table*}

\subsection{Data reductions}
The data obtained with the NOT were reduced using the {\sc iraf} package. Bias subtraction 
and flat-fielding were made in a standard manner. For the $I$--band we constructed a 
fringe-frame after removing objects from the science images. The fringe-frame 
was then subtracted from each science image, scaled to the appropriate 
background level. The images were corrected for atmospheric extinction, aligned,
 and finally combined. Calibration was done using standard stars from Landolt 
 (1992). The galaxies were corrected for galactic extinction according to 
 Schlegel, Finkbeiner \& Davis (1998). A description of data reduction and 
calibration of the MS1008--1224 images is given at the ESO web page.

Photometry was obtained using the {\sc focas} package (Jarvis \& Tyson 1981; Valdes 
1982; Valdes 1993). The detection 
limit was set to 3$\sigma$ of the sky noise, and a minimum detection area corresponding to 
the seeing-disc was used. For each object and filter, we made a catalogue listing isophotal 
magnitude, aperture magnitude, position and area. When calculating the aperture magnitude 
we first smoothed the observations to the seeing of the filter with the worst seeing of 
each cluster, and then used a constant aperture size with a radius corresponding to this seeing. 
Finally, a combined catalogue for each object was made by using the positions in the 
$R$--catalogue and matching these with the positions in the other catalogues. 

The completeness was tested with simulations where we added artificial galaxies with 
different magnitudes and radial profiles to the real images. Using the same detection 
procedures as for the real data we find that 100 per cent of the galaxies are detected down 
to $m_R~=~25$ for all our objects. Using simulations we also calculate corrections that 
should be applied when calculating total magnitudes from the observed isophotal magnitudes. 
These simulations are described in N\"{a}slund, Fransson \& Huldtgren (2000) and DFN02.  

In Table \ref{Table4} we list the number of objects in each catalogue to a limiting 
magnitude $m_R~<~25$. We also list the number of objects having aperture photometry 
in at least four filters (i.e four or five filters for Cl1601+42, and four filters 
for the other objects), three filters, as well as in two or one filter only. Objects 
identified as stars by visual inspection of the psf, are excluded.

\begin{table*}
\begin{minipage}{140mm}
\caption{Number of objects with $m_R~<~25$ in the different images, and the number of 
objects with aperture photometry in four to five filters, three filters, and one or 
two filters only. Stars are excluded.}
\begin{tabular}{lcccc}
\bf Image &\bf No. of Objects & \bf $\ge$ 4 filters & \bf 3 filters & \bf 1-2 filters \\ 
& & & \\
Cl0016+16 & 1430 & 1154 & 172 & 104\\
Cl1600+41 & 809 & 791 & 13 & 5\\
Cl1601+42 & 1199 & 1034 & 106 & 59\\
MS1008-1224 & 1665 & 1544 & 104 & 17\\
Field1 & 945 & 781 & 87 & 77\\
\end{tabular}
\label{Table4}
\end{minipage}
\end{table*}

\section{Photometric redshifts}
An extended discussion on the use of photometric redshifts applied to clusters of galaxies 
is given in DFN02. Here we present a brief summary of the technique. For every object we 
minimize the expression
\begin{equation}
\chi^2(t,z,m_{\alpha})=\sum_{i}\frac{(m_i-(T_i(t,z)+m_{\alpha}))^2}{\sigma_i^2}
\end{equation}
where $m_i$ and $\sigma_i$ are the observed magnitudes and uncertainties in
filter $i$, respectively. $T_i(t,z)$ is the magnitude in the $i$--filter of template $t$, 
redshifted to $z$. This magnitude is calculated by convolving the filter curve and the 
quantum efficiency of the detector with the galaxy template. The quantity $m_{\alpha}$ 
is a constant, which fits the apparent magnitude of the template galaxy. 

A set of ten different templates are used. We construct these by interpolations between 
the four observed galaxy templates given by Coleman, Wu \& Weedman (1980). These represent 
E, Sbc, Scd and Im galaxies. Absorption due to intergalactic \hbox{H\,{\sc i}} clouds is 
treated as in Madau (1995). We also include eight stellar templates of M--dwarfs taken 
from Gunn \& Stryker (1983).

\begin{figure}
\centerline{
\psfig{figure=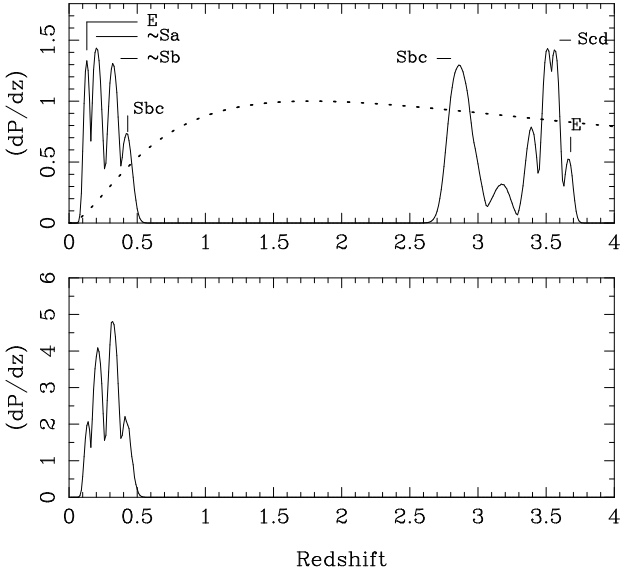,height=7.5cm}}
\caption{Chi-square probability function for the template fitting method, for 
a galaxy with known spectroscopic redshift $z~=~0.31$. At each redshift the 
probability of the galaxy template that has the minimum chi-square, i.e the 
highest probability is plotted. The top panel shows the probability function 
without considering absolute magnitudes or volume element. The dotted line 
shows the redshift dependence of the volume element. In the bottom panel 
we use a cut-off at bright magnitudes, and take the volume element into account.
For some of the peaks we plot the associated galaxy type. The distributions 
are normalised to unit area.}
\label{Figure1}
\end{figure}

A modification compared to DFN02 is that we here use a "Bayesian" approach (e.g. Kodama, 
Bell \& Bower 1999; Benitez 2000). This method allows the incorporation of pre-existing 
knowledge about the galaxies into the photometric redshift determination. This is 
illustrated in Fig. \ref{Figure1}, where we in the top panel show the probability 
distribution derived from the chi-square fit according to Eq.(1), for a galaxy with 
known spectroscopic redshift $z$ = 0.31. There are ten local minima in the chi-square 
fit, resulting in ten probability maxima, divided into two groups, one at low, and one 
at high redshifts. The best-fitting template is in this case an Sa galaxy with redshift 
$z$ = 0.20, but almost as good fits are achieved for the three other templates 
with $z~<$ 0.5, representing E to Sbc galaxies. There are also maxima with high 
probability at $z~\sim~2.8-3.5$. The degeneracy of the high- and a low-redshift groups 
arise because the Lyman-break falls approximately between the same filters at $z~\sim$ 
3.5, as the 4000-\AA~break does at $z~\sim~0.3$. 

The absolute magnitude for this galaxy at $z~\sim$ 3.5 would, however, be unrealistic, 
$M_B~\sim~-28$, whereas at $z~\sim~0.1$ and $z~\sim~0.3$ we get $M_B~\sim~-18$ and 
$M_B~\sim~-20.5$, respectively. To account for this one could weight the probability 
function with an expected LF, an approach used by Kodama et al. (1999). However, since 
we are here interested in {\em determining} the LF, we can not use this method. Instead, 
we use an exponential cut-off two magnitudes brighter than $M_B^*$, where $M_B^*$ is 
determined by a fit to the Schechter function without any weighting. The effect of 
introducing this cut-off at bright magnitudes is mainly to suppress the false peaks 
introduced by the misidentification between the Lyman--break and the 4000-\AA~break. 
An alternative would be to truncate the procedure at e.g. $z$ = 2. However, at 
$m_R~\sim$ 25, galaxies at $z~\gsim~2$ with "normal" absolute magnitudes are expected; 
a cut-off could here lead to a misidentification of these galaxies. Furthermore, the 
volume element increases with redshift (up to $z~\sim~1.8$, dotted line in the 
top panel of Fig. \ref{Figure1}). Therefore, of two
 maxima with the same probability from the chi-square fit, the higher redshift 
 should for constant comoving density, be more likely due to the larger volume.

The lower panel of Fig. \ref{Figure1} shows the probability distribution for the same 
galaxy after multiplying the result from the top panel with the probability distribution 
derived after applying an exponential cut-off at bright magnitudes and a weighting by 
the volume element. The distributions are normalised to unit area.
  
The peaks at $z~>$ 2.5 have now disappeared, and the lowest redshift peaks are suppressed 
due to the smaller volume element. The maximum probability is now for a Sb galaxy at 
$z$ = 0.31, which matches the spectroscopic redshift.

The ``Bayesian'' approach we use mostly affects the determination of redshifts for the 
cluster MS1008--1224 
at $z$ = 0.31. Liu \& Green (1998) show that when using photometric redshifts there is 
a risk of misidentification between Sbc galaxies at $z~\sim$ 0.05 and Scd/Im 
galaxies at $z~\sim$ 0.3. From our example above we also note that $\sim$ Sb galaxies 
at $z~\sim~0.3$ can be misidentified as E galaxies at $z~\lsim~0.1$.

When estimating the dispersion between spectroscopic and photometric redshifts 
for 61 galaxies in the field of 
MS1008--1224, we find $\sigma_z~\sim~0.051$, when we use the "Bayesian" approach, 
compared to $\sigma_z~\sim~0.11$ without it. The increased
dispersion is mainly due to a few galaxies achieving large errors. Excluding six
galaxies reduces the dispersion to $\sigma_z~\sim~0.058$. For the clusters at 
$z~\sim$ 0.55 the effect of including this weighting is marginal. Here we get
$\sigma_z~\sim~0.054$ with the ``Bayesian'' approach and 
$\sigma_z~\sim~0.060$ without, when calculating the dispersion between the 
photometric and spectroscopic redshifts for 38 galaxies in the field containing
Cl0016+16.

The example shown in Fig. \ref{Figure1} was picked to illustrate the 
"Bayesian" method, and has an unusual high number of probability peaks with 
similar strengths. Most galaxies have a more clearly defined primary peak.

Possible systematic effects when using photometric redshifts
are discussed in DFN02. In summary, we find a marginal increase
in the photometric redshift errors for the faintest galaxies ($m_R\gsim~24$) due to 
increased photometric errors. These errors mostly affect late type
galaxies. In numbers, we find a possible increase in redshift uncertainty by
$\Delta \sigma_z\sim~0.01$ due to systematic effects at $m_R\gsim~24$. This may
lead to an underestimate of the number of cluster galaxies in the faintest bins 
by $\sim$5\%. This result shows that systematic errors are unlikely to dominate over 
statistical errors.

\subsection{Photometric redshift catalogue}
We calculate photometric redshifts for all objects with $m_R~\le~25$ and 
aperture photometry in at least three filters, which corresponds to 92--99 per 
cent of the total number of objects to this limit (Table \ref{Table4}). The 
accuracy of the photometric redshifts is estimated by comparing our result 
with available spectroscopic redshifts. In Table \ref{Table5} we list the 
number of available redshifts and the rms deviation between the photometric 
and spectroscopic redshifts, $\sigma_z$. The galaxies used for calculating the rms have
photometry in all filters. To estimate the increase in errors for objects 
only having photometry in three filters, we also calculate the rms deviation 
after excluding the filter with the largest error for each galaxy, $\sigma^{(3)}_z$.
\begin{table*}
\caption{N is the Number of galaxies with spectroscopic redshifts used to 
determine the dispersion between photometric and spectroscopic redshifts, 
$\sigma_z$ is the dispersion using information from all filters, while 
$\sigma^{(3)}_z$ is the dispersion using only three filters. Also given is the 
resulting number of clusters galaxies within $z_{Cl}\pm1.5\sigma_z$ for the 
total areas and inside a common radius of 1 Mpc. We also list the density of 
background galaxies within each redshift range. The limiting magnitude is 
$M_B=-17.7$, corresponding to $m_R~\sim~25$ at $z=0.54$ and $m_R~\sim~23.4$ 
at $z=0.31$.}
\begin{tabular}{llllllll}
\bf Cluster & $z_{Cl}$ & N & $\sigma_z$ & $\sigma^{(3)}_z$ & \multicolumn{2}{l}{Cluster galaxies} & Background$^a$\\
& & & & & Total area & $R~<~1~{\rm Mpc}$ & (${\rm Mpc}^{-2}$)\\
Cl0016+16 & 0.546 &38$^b$ & 0.054 & 0.080$^c$ & $463\pm32$ & $248\pm19$ & 12.7$\pm2.1$\\
Cl1600+41 & 0.540 &$^d$ & 0.065$^d$ & 0.10$^d$ & $~36\pm24$ & $~26\pm12$ & 14.6$\pm2.4$\\
Cl1601+42 & 0.539 &78$^b$ & 0.076 & 0.12$^e$ & $332\pm35$ & $154\pm18$ & 18.8$\pm2.9$\\
MS1008-1224 & 0.306 &61$^f$ & 0.051 & 0.067 & $221\pm20$ & $173\pm16$ & 11.6$\pm2.3$\\
MS1008-1224$^g$   &  &  &  &  & $319\pm26$ & $240\pm21$ & 20.6$\pm3.5$\\
\hline 
\multicolumn{8}{l}{Notes:}\\
\multicolumn{8}{l}{$a$) Measured within the redshift range defining the different clusters, 
i.e. $z_{Cl}\pm1.5\sigma_z$.}\\
\multicolumn{8}{l}{$b$) Dressler et al. 1999.}\\
\multicolumn{8}{l}{$c$) Excluding two outliers with large errors.}\\
\multicolumn{8}{l}{$d$) No spectroscopic redshifts available, except for central 
galaxy. We assume dispersions}\\
\multicolumn{8}{l}{equal to the mean of the dispersions of the two clusters at similar redshift.}\\
\multicolumn{8}{l}{$e$) Excluding five outliers.}\\
\multicolumn{8}{l}{$f$) Yee et al. 1998.}\\
\multicolumn{8}{l}{$g$) Number of galaxies to $M_B=-16.2$.}\\
\end{tabular}
\label{Table5}
\end{table*}

\subsection{Comment on background counts}
To estimate the background contamination we use $BVRI$ photometry of our blank field. 
In DFN02 we used $UBVRI$ photometry of the ESO Imaging Survey (EIS) deep field published 
by da Costa et 
al. (1998). Number counts to $R$ = 25 for Field1 and the EIS field yield 
983 $\pm$ 133 and 845 $\pm$ 120 galaxies, respectively, where the fields are normalised to the 
same size as the image of Cl0016+16 (32.5 sq. arcmin). Errors represent 1$\sigma$ and include Poisson 
statistics and field-to-field variance (see DFN02 for a discussion). The resulting 
counts are therefore consistent within the errors. 

When determining the background counts within the cluster redshift range, we find a 
larger deviation between the two fields. In the redshift range 
$z_{Cl0016}~\pm~1.5\sigma_z$, we find 121 $\pm$ 20 galaxies for Field1, and 166 $\pm$ 
26 galaxies for the EIS field to $R$ = 25. These results are marginally within errors, 
but it is likely that there are systematic effects responsible for some of the 
off-set. 
In particular, the aperture magnitudes are determined differently. For our blank field 
we smooth the images to the same seeing before calculating the colours, whereas the 
aperture magnitudes given for the EIS field have a varying seeing in the different 
bands. This can affect the colours, especially for small objects that do not cover 
the whole aperture. 

To check the effect of this, we calculate photometric redshifts for the field 
containing Cl0016+16 without smoothing the images to the same seeing. Comparing 
with the 38 objects 
that have spectroscopic redshifts we find an rms deviation $\sigma_z~=~0.078$, which 
is considerably larger than what we found after smoothing, $\sigma_z~=~0.054$. Therefore, 
the photometric redshifts calculated for the EIS catalogue are likely to be less 
accurate. 

Finally, systematic errors could be introduced by 
the uncertainty in the zero-point magnitudes and the use of different software when 
doing the photometry, i.e the use of {\sc focas} for Field1, and SExtractor for the 
EIS field. 

To minimize the risk of introducing systematic errors, we use the photometry from our 
blank field in this analysis, since this is derived in the same way as the photometry 
of the cluster images. The use of a different background field does, however, not affect 
the result in DFN02 on Cl1601+42 more than marginally, i.e. any differences are 
within the quoted errors. This illustrates the advantage of the photometric method for 
examining high-$z$ clusters; the relatively small corrections for background 
contamination makes the results less dependent on the background. 
\subsection{Cluster membership and galaxy classification}
For selecting cluster members we use a redshift range $\pm$ 1.5$\sigma_z$ around the 
mean redshift of the cluster, with values of $\sigma_z$ given in Table \ref{Table5}. 
To account for the increase in dispersion for the galaxies only detected in three 
filters (Table \ref{Table3}), we use the corresponding $\sigma^{(3)}_z$ for these galaxies.
Note, however, that the information given by a non-detection (i.e., that the 
magnitude of the galaxy is fainter than the magnitude limit of the observations) is taken into
account in the template fitting procedure. Assuming that 
the errors are Gaussian, the 1.5$\sigma_z$ cut should include $\sim$ 86 per cent of 
the actual number of cluster galaxies to our limit. Taking into account the small 
fraction of the galaxies observed in three filters only that may result in outliers 
with large errors, we estimate that we include $\sim 80-85$ per cent of the total number of 
cluster galaxies when applying the 1.5$\sigma$ cut. For a discussion on completeness 
and contamination when selecting cluster members using photometric redshift, see 
Brunner \& Lubin (2000). 

In this study we divide the galaxies into late-types and early-types, based on the 
spectral type determined by the photometric colours of the galaxies. The division is 
made half way between the E and Sbc templates from Coleman et al. (1980), approximately 
corresponding to a rest-frame colour $B-V~=~0.8$. The first category consists of 
early-type galaxies with red colours, which are mostly ellipticals and lenticulars, 
but also include passive spirals with red colours. The second category includes 
late-type spirals and irregulars, with a possible inclusion of blue elliptical 
systems.

\section{Results}
\begin{figure*}
\centerline{
\psfig{figure=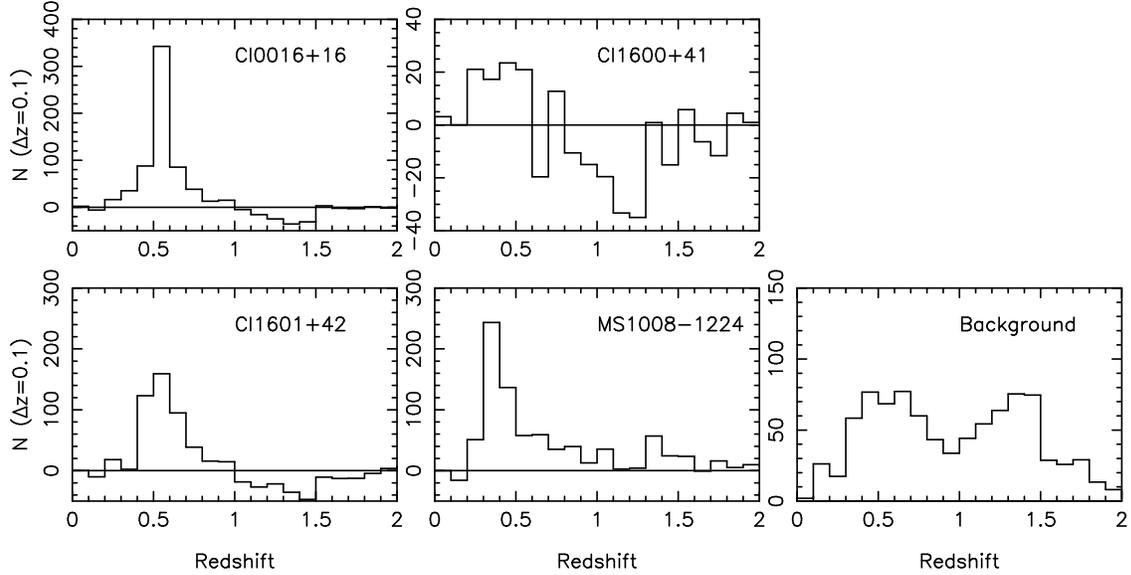,height=7.6cm}}
\caption{The left and the middle panels show the distribution of photometric redshifts 
in the cluster images after subtracting background galaxies. The right-hand panel shows the
redshift distribution of the background field. Note the different scales on the $y$-axis.
The peak in the background distribution at $z~\sim~1.4$ is 
most likely caused by the lack of infrared photometry. All distributions above $z~\sim~1$ 
should therefore be viewed by caution.}
\label{Figure2}
\end{figure*}

The distribution of photometric redshifts in the cluster fields, after subtracting 
background galaxies, is shown in the left and middle panels in 
Fig. \ref{Figure2}. The right-hand panel shows the redshift distribution of the 
background field. The location of the clusters 
is apparent for all clusters except Cl1600+41. We comment on this cluster in next 
subsection.

As a measure of the richness of the clusters we give in Table \ref{Table5} the number of 
cluster galaxies with $M_B~<~-17.7$ for the total areas (see Table \ref{Table3}), as well 
as inside a common radius of 1 Mpc. We also list the density of background galaxies 
within the redshift range adopted for the different clusters. For the lower-redshift 
cluster, MS1008--1224, we also give numbers for $M_B~<~-16.2$.

\begin{figure}
\centerline{
\psfig{figure=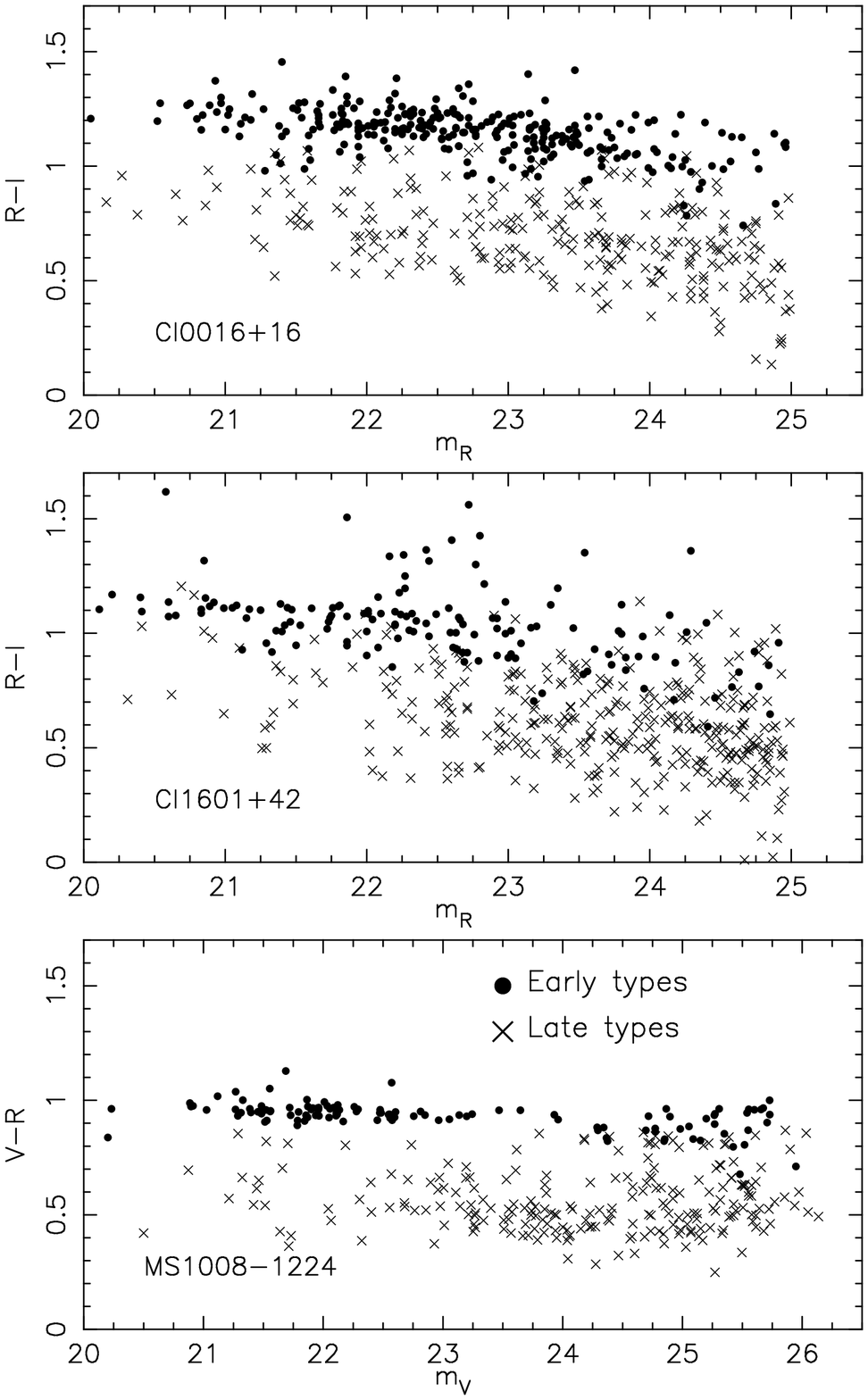,height=13cm}}
\caption{Obsereved colour--magnitude diagrams for Cl0016+16 (top), Cl1601+42 (middle) and
MS1008--1224 (bottom). Galaxies are separated into early-types and late-types. At the redshifts
of the clusters, the observed colours approximately match rest-frame $B-V$ colour.}
\label{Figure4}
\end{figure}

In the left-hand panel of Fig. \ref{Figure4}, we show the surface density of galaxies 
with $M_B~<~-17.7$ for 
the four clusters as a function of radius. The horizontal solid line marks the surface 
density of the background field in the redshift range of Cl1600+41. Error bars and dashed 
lines represent 1$\sigma$ errors. The right-hand panel of Fig. \ref{Figure4} shows the 
projected fraction of early-type galaxies for the four clusters. The horizontal line 
marks the early-type fraction of the background field at $z~\sim~0.55$, which 
is also similar to 
what we find at $z~\sim~0.31$. All clusters have an early-type fraction clearly above 
the field value in the core region. In Cl0016+16 this fraction is $\sim$85 per cent in 
the core. This decreases with radius, but stays above 50 per cent for the whole area covered.
The early-type fractions in Cl1601+42 and MS1008--1224 is $\sim$60 per cent in the core, and show 
a decrease which approaches the 
field value at large radius, even though the deviation from the field is still significant 
at the outermost points. For Cl1600+41, only the innermost point has a late type fraction 
deviating from the field. From both figures it is clear that we do not reach the 
field in any of the clusters, except Cl1600+41.

\begin{figure*}
\centerline{
\psfig{figure=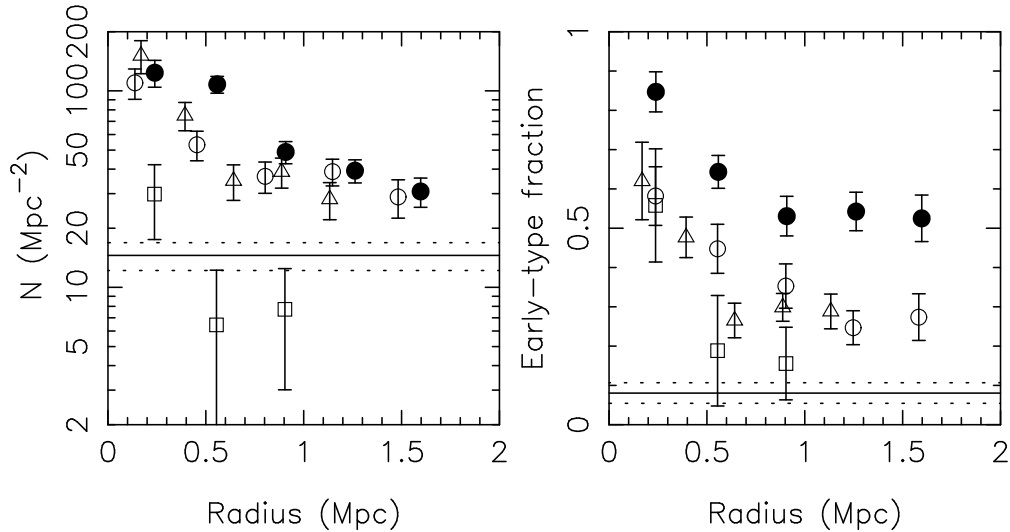,height=7cm}}
\caption{Left-hand panel: Projected surface density as function of radius for Cl0016+16 
(filled circles), Cl1600+41 (squares), Cl1601+42 (open circles) and MS1008--1224 
(triangles). The horizontal line shows the surface density of the background field in 
the redshift range of Cl1600+41. Right-hand panel: Early-type fraction as function of radius 
for the four clusters. The horizontal line shows the early-type fraction of the field 
at $z~=~0.54$. The limiting magnitude is $M_B~=~-17.7$. Error bars and the dashed lines 
represent 1$\sigma$ errors.}
\label{Figure3}
\end{figure*}

In Fig. \ref{Figure3}, we show the observed colour--magnitude (CM) diagram for all 
galaxies within the cluster redshift ranges for the three
rich clusters, with different symbols representing early-type and late-type galaxies. At
the redshifts of the clusters, the plotted colours approximately represent rest-frame $B-V$.
From the figure, it is evident that the early-type galaxies populate a fairly narrow sequence 
in the CM diagram, which is clearly redder than the late-type galaxies. Also, the relative
fraction of early-type galaxies decreases at fainter magnitudes. Note, however, that the spectral 
type is not independent of e.g., the rest-frame $B-V$~colour, since colours are used when determining 
the photometric redshift, as well as spectral type.

\subsection{The poor cluster Cl1600+41}
From the results above it is clear that Cl1600+41 is very poor, with only $36\pm24$ 
cluster galaxies with $m_R~<~25$. Fig. \ref{Figure4} also indicates that if at all 
significant, the radial extent of the cluster is less than $\sim$ 0.5 Mpc. At larger 
radii the surface density is marginally below the field value, while the late type 
fraction is consistent with the late type fraction of the field.

The most significant indication of a cluster comes from the colour distribution, where the 
innermost point shows a clear excess of red galaxies (Fig. \ref{Figure4}, right-hand panel). Also, the 
magnitude and colour of the central galaxy is similar to the brightest cluster galaxy 
(BCG) in the other clusters. Furthermore, a visual inspection of a three-colour image 
of Cl1600+41 shows this bright red central galaxy to be surrounded by a number of red 
and blue galaxies, clearly indicating the presence of a cluster.

If we calculate the number of cluster galaxies in Cl1600+41 by the standard method of 
subtracting galaxies in a blank field from the cluster image in a {\it single} band, 
we get a total of 9 $\pm~127$ cluster galaxies to $m_R~<~25$. The large error is due 
to field-to-field variations, which dominate in the subtraction method. It is obvious that the 
subtraction method would not result in any cluster detection when using our blank field 
as a reference for the background counts.

With only 36 selected cluster members within the chosen redshift range, the background 
counts dominate over the cluster galaxies, i.e. only 22 per cent of the galaxies are 
expected to be cluster galaxies, as compared to 78, 66 and 80 per cent, for Cl0016+16, 
Cl1601+42 and MS1008--1224. It is clear that this cluster is close to the detection limit. 
The large contamination in Cl1600+41 makes it impossible to determine either the internal 
properties of the cluster with any significance, or the cluster LF. Therefore, we do 
not include Cl1600+41 when comparing the LF and the Butcher--Oemler (BO) effect between the clusters.

\subsection{The cluster luminosity functions}
When calculating the LF we divide the galaxies within the cluster redshift range into 
magnitude bins with $\Delta m~=~0.5$ to $M_B~=-17.7$ for the two clusters at $z~\sim~0.55$,
and to $M_B~=-16.2$ for MS1008--1224 at $z~\sim~0.31$. K-corrections are determined for 
each galaxy using the best-fitting template from the photometric redshift calculations. 
For each magnitude bin we subtract background galaxies within the redshift range of the 
different clusters.

\begin{figure*}
\centerline{
\psfig{figure=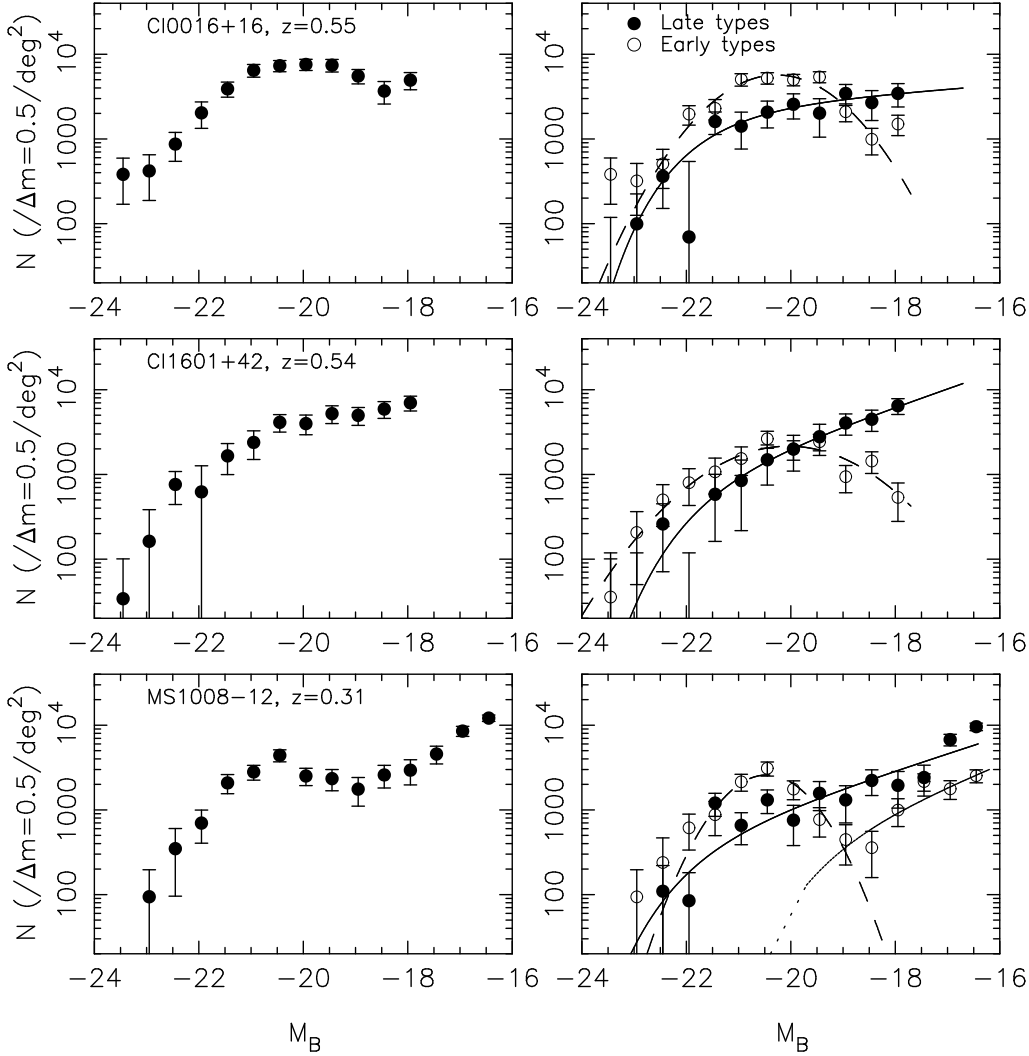,height=14cm}}
\caption{Luminosity functions in the rest-frame $B$--band for Cl0016+16, Cl1601+42 and 
MS1008--1224. The left-hand panels show the total LFs, while in the right-hand panels the LFs are 
divided into early-type (open circles) and late-type (filled circles) galaxies. 
Error-bars include Poissonian uncertainties and field-to-field variance. For MS1008--1224 
we show separate Schechter fits for the early-type and late-type populations.}
\label{Figure5}
\end{figure*}

The left-hand panels in Fig. \ref{Figure5} show the resulting LFs for the three clusters. 
As a first step we fit the total LFs to the usual Schechter function 
(Schechter 1976),
\begin{equation}
\Phi (M)\propto e^{-10^{0.4(M^*-M)}}10^{-0.4(\alpha +1)M},
\end{equation}
where $M^*$ is the characteristic magnitude, representing the turnoff at the bright end 
of the LF profile, and $\alpha$ is the slope at the faint end of the LF. As an alternative 
to $\alpha$, we also calculate the slope of a straight line fit to the five faintest 
bins in each LF, according to
\begin{equation}
\Phi_f (M)\propto 10^{-0.4(\alpha_f +1)M}
\end{equation}
(Trentham 1998a; DFN02). The parameter $\alpha_f$ gives a better representation of the 
faint-end of the LF, since 
it is not affected by the coupling between $M^*$ and $\alpha$ in the Schechter function. 
In Table \ref{Table6} we list $M^*$, $\alpha$ and $\alpha_f$ for the clusters. Parameters 
are given both for the total cluster populations, and for the late-type population in 
each cluster.

\begin{table*}
\caption{Characteristic magnitude, $M_B^*$ and faint-end slope, $\alpha$ derived from 
the Schechter function fit to the rest-frame $B$--band LFs for the total population and 
for late-type galaxies only. $\alpha_f$ is the slope of a straight-line fit to the five 
faintest-magnitude bins in each cluster. Limiting magnitude is $M_B~=~-17.7$. For 
MS1008--1224, we also give results to $M_B~=~-16.2$. For the early-type population, 
$M^G_B$ is the peak magnitude and $\sigma^G$ is the width of a Gaussian fit to 
$M_B~\le~-18.0$.}
\begin{tabular}{lcccccccc}
 \multicolumn{1}{l}{\bf Cluster} & \multicolumn{3}{l}{\bf All galaxies} & \multicolumn{3}{l}{\bf Late-types} & {\bf Early-types} & \multicolumn{1}{l}{}\\
& $M_B^*$ & $\alpha$ & $\alpha_f$ & $M_B^*$ & $\alpha$ & $\alpha_f$ & $M^G_B$ & $\sigma^G$\\
Cl0016+16 & -21.07 $\pm0.24$ & -0.55 $\pm0.10$ & -0.65 $\pm0.10$ & -21.75 $\pm0.50$ & -1.12 $\pm0.17$ & -1.17 $\pm0.18$ & -20.27 $\pm0.08$ & 1.01 $\pm0.05$\\
Cl1601+42 & -21.87 $\pm0.36$ & -1.25 $\pm0.10$ & -1.33 $\pm0.14$ & -21.80 $\pm0.53$ & -1.53 $\pm0.27$ & -1.65 $\pm0.19$ & -20.05 $\pm0.17$ & $1.30\pm0.10$\\
MS1008-1224 & -21.06 $\pm0.17$ & -0.72 $\pm0.26$ & -0.99 $\pm0.15$ & -21.05 $\pm0.58$ & -1.11 $\pm0.33$ & -1.61 $\pm0.24$ & -20.46 $\pm0.10$ & $ 0.73\pm0.06$\\
MS1008-1224$^a$ & -21.48 $\pm0.17$ & -1.46 $\pm0.09$ & -1.92 $\pm0.13$ & -22.00 $\pm0.37$ & -1.50 $\pm0.17$ & -1.89 $\pm0.16$& &\\
\hline
\multicolumn{9}{l}{Note:}\\
\multicolumn{9}{l}{a) Limiting magnitude $M_B~=~-16.2$.}
\end{tabular}
\label{Table6}
\end{table*}

Fig. \ref{Figure5} shows that the shapes of the LFs of the three clusters differ
 substantially. To 
further understand these differences, we plot the LFs divided into early-type and 
late-type galaxies separately in the right-hand panels of Fig. \ref{Figure5}. For all 
three clusters, the early-type galaxies with $M_B~\lsim~-18$ have a Gaussian LF, 
peaking at $M_B^G~\sim~-20$ (Table \ref{Table6}). We note that the faint early-type 
population in MS1008--1224 shows an increase below $M_B~=~-18$. We return to this point 
in Section 5.

The late-type galaxies are better fitted by Schechter functions. However, the 
relative normalisations of the early-type and late-type LFs vary between the 
different clusters. In Section 5 we 
discuss these differences further.   
\subsection{The Butcher--Oemler effect}
We calculate the blue fraction, $f_B$, using the definition in BO84. For each
cluster we select galaxies with $M_V~<~-20$ within a circular radius, $R_{30}$,
containing 30 per cent of the cluster galaxies. Now, $f_B$ is the fraction of these
galaxies that are at least 0.2 mag bluer than the colour--magnitude relation for
the cluster. To account for cluster galaxies outside our field-of-view, we fit
the cluster radial profile with a singular isothermal sphere when calculating
$R_{30}$.   

In order to compare results, we adopt in this section only the cosmology used by BO84 ($q_0~=~0.1$). 
The difference from BO84 is that we here select cluster members 
with the photometric redshift method. This increases the accuracy since background 
subtraction is reduced. The blue fractions derived are given in Table \ref{Table7}. 

For MS1008--1224, we find a value of $f_B$ that matches the straight-line fit of the blue 
fraction as a function of redshift presented in BO84. For Cl1601+42 and Cl0016+16 the 
values of $f_B$ are located below this fit. The scatter in the fit is, however, 
large, and at least for Cl1601+42 this deviation is hardly significant. Cl0016+16 
is included in the BO84 sample, and they also note that this cluster is exceptional in 
the sense that it has a fraction of blue galaxies that is more similar to local clusters 
than to clusters at $z~>~0.3$. The blue fraction found by BO84 is $f_B$ = 0.02 $\pm$ 0.07, 
which is consistent with our estimate.
\subsection{Brightest cluster galaxies}
In Table \ref{Table7} we list the absolute $B$ magnitude and rest-frame $B-V$ colour for 
the brightest cluster galaxy (BCG) in the observed clusters, including the poor cluster 
Cl1600+41. Even though our sample is small, the results are consistent with the magnitude 
of the BCG being independent of the richness of the cluster, as shown by e.g. Sandage 
(1976) and Postman \& Lauer (1995). In particular, we note that the BCG of the extremely 
poor cluster Cl1600+41 is only marginally fainter than the rest, demonstrating that even 
in this environment a luminous elliptical can be formed.
\begin{table*}
\begin{minipage}{140mm}
\caption{The blue fraction, $f_B$, as defined by BO84, the absolute $B$ magnitudes and 
rest-frame $B-V$ colour for the brightest cluster galaxy in our cluster sample and the 
early-type fraction at two limiting magnitudes.}
\begin{tabular}{lcccc}
\bf Cluster & $z$ & $f_B$ &\bf $M_B$ & \bf $B-V$ \\ 
Cl0016+16 & 0.546 & $0.04~\pm0.02$ & -23.52 $\pm0.02$ & 1.00 $\pm0.03$ \\
Cl1600+41 & 0.540 & - & -23.36 $\pm0.03$ & 0.92 $\pm0.05$ \\
Cl1601+42 & 0.539 & $0.15~\pm0.06$ & -23.50 $\pm0.04$ & 0.93 $\pm0.07$ \\
MS1008-1224 & 0.306 & $0.13~\pm0.06$& -23.65 $\pm0.05$ & 1.15 $\pm0.07$ \\
\end{tabular}
\label{Table7}
\end{minipage}
\end{table*}

\section{Discussion}
\subsection{Variations in the cluster luminosity function}
The different shapes of the LFs in our sample suggest that there is no universal 
form of the total cluster LF at $z~\gsim~0.3$ (Fig. \ref{Figure5}). 
To quantify this we construct an average LF from the three individual
best-fitting LFs. We then calculate the reduced chi-square when fitting each 
LF to the average LF. We also calculate the reduced chi-square
when fitting the individual LFs to a Gaussian+Schechter function. 
Results are shown in Table 8. 
The chi-squares are significantly higher for the fits to the average LF, 
supporting our claim for a non-universal LF.
We also find that 
a single Schechter function gives a poor representation of the cluster LF, which agrees 
with e.g. Driver et al. (1994) and Wilson et al. (1997). This is especially evident for 
MS1008--1224, where the best Schechter fit to $M_B~<~-17.7$ yields a faint-end 
slope that is decreasing, while the data-points clearly show an increase 
in the number of faint galaxies with $M_B~>~-19$. As in the case
for Virgo, a better representation is given by the sum of a Gaussian
and Schechter function. 
The reduced chi-square for for a Gaussian+Schechter function is $\chi^2/\nu~=~0.98$ , while
a single Schechter function yields $\chi^2/\nu~=~5.8$.

From the separate late-type and early-type LFs (right-hand panels of Fig. 
\ref{Figure5}), it is clear that the relative abundance of the early-type and the 
late-type populations determines the over-all shape of the LF. 
In Cl1601+42 late-type 
galaxies dominate at $M_B~\gsim~-20$. This results in a total LF that increases over 
the whole magnitude range, and has a steep faint-end slope. MS1008--1224 is similar to 
Cl1601+42 in that it contains a numerous population of faint 
late-type galaxies, which dominates the total LF at faint magnitudes. The magnitude 
where this cross-over takes place is, however, slightly fainter in MS1008--1224 
($M_B~\sim~-19.5$) compared to Cl1601+42 ($M_B~\sim~-20$), which leads to a total LF 
where both an intermediate-magnitude Gaussian part, as well as a steep faint-end slope 
of late-type galaxies are distinguishable. Note also that MS1008--1224 has a population 
of faint {\it early-type} galaxies, which adds to this.

In Cl0016+16 there are relatively few faint late-type galaxies, compared to the number 
of early-type galaxies at $-21~\lsim~M_B~\lsim~-19$. This leads to a total LF that 
has a Gaussian shape at intermediate magnitudes, and only at the faintest bin is there
an indication of a rise. 

As discussed in DFN02, the differences between the shapes of the cluster LFs can be 
explained by the fact that clusters in the hierarchical clustering scenario at high 
redshift have a larger fraction of newly accreted galaxies with ongoing star 
formation. A general increase in the star-formation rate in field galaxies with 
redshift (Diaferio et al. 2001), also contributes to an increase of star-forming 
galaxies in clusters at high $z$. 

After accretion, and a possible period of enhanced star formation, the galaxies 
could have their gaseous envelopes removed by tides or ram pressure stripping, 
leading to a fading over time-scales of Gyrs, as the remaining gas reservoir is 
exhausted ('strangulation') (Balogh et al. 2000; 
Diaferio et al. 2001). Such fading of the faint blue galaxies is discussed by 
Wilson et al. (1997), who find that a fading of dwarf irregulars by 
$\sim~3$ mag can explain the difference between the LF in clusters $z~\sim~0.2$
and the local Virgo cluster. Further, Conselice et al. (2001) show that spirals 
accreted at high $z$ can be transformed into the faint dE population seen in 
local clusters by "galaxy harassment" (Moore et al. 1998). 

It is of obvious interest to compare the LFs of the different populations of these 
clusters with those of nearby clusters like Coma and Virgo. We then have the 
paradoxical situation that studies of nearby cluster are limited to one, two or at 
most three colours, or alternatively to a morphological separation of different classes. 
A direct comparison with our spectral classification is therefore difficult. As a first 
step we therefore compare the {\it total} LF of our clusters with that of Coma at $z~=~0.02$ 
(Trentham 1998a) and Virgo $z~=~0.003$ (Trentham \& Hodgkin 2002).

In Fig. \ref{Figure6} we plot the rest-frame $B$--band LF for Cl1601+42 (open 
circles), Cl0016+16 (filled circles), and MS1008--1224 (triangles) together with the LF 
for the rich Coma cluster (squares) and the poor Virgo cluster (diamonds). For 
clarity, the LFs have arbitrary 
off-sets in the $y$-direction. As suggested by our discussion of the 
individual LFs, we represent the total LF as a sum of a Gaussian and a Schechter 
function. 
Because the bright and faint populations have distinctly different
properties, we discuss them below separately.
\begin{table}
\caption{Reduced chi-squares when comparing the LFs of Cl0016+16, Cl1601+42 and MS1008--1224 to
the best-fitting Gaussian+Schechter function, as well as an LF which is an average of the three
individual LFs.}
\begin{tabular}{lcc}
\bf Cluster & $\chi^2/\nu$  & $\chi^2/\nu$ \\
~ & (individual fit)  & (average fit) \\
Cl0016+16 & 0.45 & 1.25\\
Cl1601+42 & 0.46 & 1.49\\
MS1008-1224 & 0.98 & 1.46\\
\end{tabular}
\label{Table8}
\end{table}
\subsubsection{The bright population}
For Coma the Gaussian peaks at $M_B^G~=-19.3~\pm0.1$, indicating a fading of the 
bright population by $\Delta M\sim~1$ mag, compared to the high-redshift clusters 
(Table \ref{Table6}). The total LF of the Virgo cluster 
only has a marginal Gaussian peak. However, the population of 
morphologically classified elliptical galaxies, as well as giant spirals, has a 
clear Gaussian shape 
peaking at $M_B^G~\sim~-19$ (Binggeli et al. 1988; Ferguson \& Sandage 
1991), supporting a fading of this population similar to that suggested for Coma.

These results are consistent with Smail et al. (1997), who find 
that the characteristic magnitude $M_V^*$ for early-type galaxies fade by 
$\Delta M\sim~0.7$ mag between $z~=~0.54$ and $z~=~0$. (We 
have here converted the results in Smail et al. to our adopted cosmology.)
Smail et al. find a weaker 
trend for bright late-type galaxies, which fade by $\Delta M\lsim~0.4$ mag. An evolution 
is also found by Kodama \& Bower (2001), who estimate that the bright blue galaxy population 
fades by $\Delta M~\sim~1$ mag between $z~\sim~0.4$ and $ z~=~0$.
De Propris et al. (1999) investigate the $K$--band luminosity evolution of the bright 
galaxy population in clusters at $0.1~<~z~<~1$, and find that $K^*(z)$ is consistent 
with passive luminosity evolution. 
We have compared our luminosity evolution with
stellar synthesis models from Bruzual \& Charlot (2003). We model the Gaussian
population with an early-type population characterized by a single burst stellar
population of age 8.0 Gyr and a Salpeter IMF. This combination results in colours
that match the early-type spectral energy distribution in Coleman et al. (1980). 
For passive evolution,
we find that this population
should fade by $\sim~0.7$~mag since $z\sim~0.55$. 
This is slightly less than the $\sim~1$~mag evolution that our results suggests. 
This difference will diminish if we either assume a Scalo IMF,
or that the Gaussian population includes a fraction of galaxies with residual
star formation.
\subsubsection{The faint population}
In DFN02 we argued that the different shapes of the LF at $M_B~\gsim~-20$ between
Coma and Cl1601+42 can be understood as a result of the dynamically younger age 
of the latter 
cluster. The steep blue end of the LF in this cluster should then consist of 
recently accreted 
field galaxies, which is supported by the fact that this part of the LF is almost 
exclusively made up of late-type galaxies. The rapidly increasing fraction of 
late-type galaxies in the outer parts of the cluster is consistent with this accretion 
scenario, and is also reflected in the rising blue fraction at large radii in this 
cluster, as shown in Section 5.4. A fading of the blue star forming galaxies in Cl1601+42 
by $\sim~2$ mag, would transform the LF of Cl1601+42 into a LF similar to Coma.

\begin{figure}
\centerline{
\psfig{figure=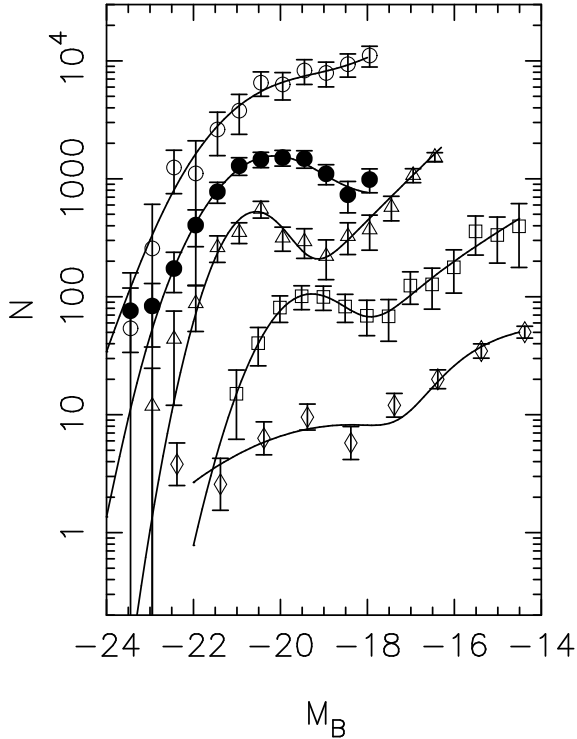,height=9.8cm}}
\caption{Luminosity function in the rest-frame $B$--band for Cl1601+42 at $z~=~0.54$ 
(open circles), Cl0016+16 at $z~=~0.55$ (filled circles),  MS1008--1224 at $z~=~0.31$ 
(triangles), Coma at $z~=~0.02$ (squares) and Virgo at $z~=~0.003$ (diamonds). 
Data for Coma and Virgo are taken from Trentham (1998a) and Trentham \& Hodgkin (2002), 
respectively. The LFs are arbitrary off-set in the $y$-direction.}
\label{Figure6}
\end{figure}

The steepening of the faint-end in MS1008--1224 is clearly shown in Fig. 
\ref{Figure6}. The slope is somewhat steeper than for Coma, and is shifted 
towards brighter magnitudes. From Fig. \ref{Figure5} we see that the faint-end of 
MS1008--1224 is made up of both blue and red galaxies. The slope of the faint blue 
population is similar to the blue population in Cl1601+42, and a fading by $\sim~1$ mag 
of this population would result in a LF similar to the one in MS1008--1224. 
A subsequent fading by an additional magnitude will make both Cl1601+42 and MS1008--1224 
similar to Coma. The fading of the late-type population between $z~=0.5$ and $z~=0$ 
for these two clusters could therefore be described by a relation $\Delta m~\simeq~-4z$ mag.

\subsubsection{The red cluster Cl0016+16}
The LF of Cl0016+16 at $z~=0.55$ differs clearly from Cl1601+42 at the same redshift. 
The LF of the former is similar to Coma in the plateau region, having a Gaussian shape,
but shifted to brighter magnitudes by $\sim~1$ mag. From Fig. 
\ref{Figure5} we see that the relatively low abundance of late-type galaxies in 
Cl0016+16, as compared to Cl1601+42, is responsible for the different shapes of the 
LF at intermediate magnitudes. 

Previous investigations of Cl0016+16 show that the bright galaxy population is 
dominated by red early-type galaxies. Koo (1981) found that this cluster has an 
unusually high fraction of red galaxies compared to other high-redshift clusters, and 
suggested that the star formation must have ended a few Gyrs before the observed epoch.
From the absence of blue galaxies and the small scatter in the colours of the red 
population, Smail et al. (1995) conclude that the cluster is old, despite its high 
redshift. Of the ten clusters in the MORPHS 
sample (Smail et al. 1997) with morphological classification from HST in the redshift 
range $0.37~<~z~<~0.56$, Cl0016+16 has the lowest fraction of spiral galaxies, 
$f_{sp}$ = 21 per cent, compared to a mean of $f_{sp}$ = 44$\pm7$ per cent for the remaining nine 
clusters. 

The low spiral fraction is, not unexpectedly, related to the low blue fraction found in Cl0016+16, which 
indicates that the cluster does not follow the general blueing described by the 
BO-effect. This may be explained by a 
higher dynamical age for Cl0016+16, which is supported by the high X-ray luminosity and 
velocity dispersion (Table \ref{Table3}), indicating that Cl0016+16 is more relaxed 
than Cl1601+42. 

Cl0016+16 has a very
high fraction of post-starburst galaxies. These galaxies have no current star formation, but 
were forming stars $\sim$1--2 Gyr before we observe them. Poggianti et al. 
(1999) found that 32$\pm9$ per cent of the galaxies brighter than $M_V\sim-20.5$~in Cl0016+16 are spectroscopically consistent 
with this type of galaxies. This suggests that a large fraction of the galaxies in 
Cl0016+16 has experienced star formation since $z~\sim~0.8$, but have subsequently faded
to become red at $z~=~0.54$. Post-starburst galaxies are less frequent in local
clusters (Poggianti et al. 1999), and although not actively forming
stars, Cl0016+16 is therefore not similar to a local 
cluster placed at high redshift. This is also supported by the brighter early-type 
population in this cluster compared to local clusters.  

The difference between rich clusters at high and low redshift can qualitatively be 
explained within the hierarchical clustering scenario. Kauffmann (1995) shows that 
high-$z$ rich clusters assemble over a shorter time interval than low-$z$
clusters of similar richness. In general, this naturally leads to a higher fraction of blue galaxies 
in high-$z$ clusters. The properties of Cl0016+16 can then be understood if it formed during a
short time interval at $z~\gsim~0.8$, leaving a high fraction of post-starburst galaxies. 
This suggests that the cluster was blue at $z~\sim~0.8$, while the red colours at $z~=~0.55$
indicates a very low infall of field galaxies during the last $\sim~1$ Gyr before it is 
observed. 

It is possible that the special environment of Cl0016+16 could have an 
influence on the blue fraction and the cluster LF. Connolly et al. (1996) and Hughes \& 
Birkinshaw (1998) 
provide strong evidence that Cl0016+16 is part of a supercluster structure, with two 
associated clusters at similar redshift with projected distances of 5 and 13 Mpc. 
The formation of this giant structure could have depleted the number of field galaxies 
surrounding Cl0016+16, and therefore the accretion rate of late type galaxies.
\subsubsection{A universal cluster LF?}
To summarise our discussion, we find no evidence for a universal shape of the total
cluster LF, in agreement with e.g., Binggeli et al. (1988) and Driver, Couch \& Phillipps 
(1998). The LFs for the spectroscopic early-type and late-type galaxies separately have, 
however, similar shape in the different clusters, but with varying
relative strengths. 
This suggests an universality for type-specific LFs. This is analogous to the claim of
type-specific LFs for different morphological types by Binggeli et al. (1988)
and Andreon (1998). A comparison with Binggeli et al. (1988) is especially interesting,
since they find that the bright part of the LF mainly consists of elliptical galaxies
with a Gaussian LF, and that the steep faint-end consists of irregulars and dEs.
This suggests, as expected, that the early-type population mainly consists of ellipticals, while
many faint late-type galaxies are irregulars.
To explain the different shapes of the high-$z$ cluster 
LFs, we argue that these reflect different dynamical states of the clusters. 
As the clusters get dynamically 
older, we expect the late-type population to fade relative to the early-type 
population, and LFs to become more similar. 
\subsection{The dwarf population in MS1008--1224}
\begin{figure}
\centerline{
\psfig{figure=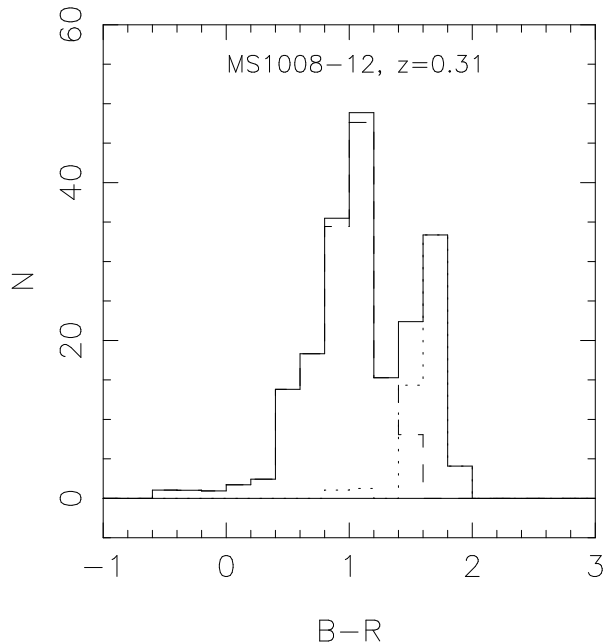,height=8.5cm}}
\caption{The rest-frame $B-R$ colour distribution for galaxies in MS1008--1224 
with $-17.7~<~M_B~<~-16.2$. 
The solid line shows the total number of galaxies, while the dotted and dashed lines show 
the distribution of early-type and late-type populations, as classified by multi-band 
photometry, separately.}
\label{Figure7}
\end{figure}

The faint limit in absolute magnitude we reach for MS1008--1224 at $z~=~0.31$ allows
us to study the dwarf population $\sim~1.5$ mag deeper in this cluster 
compared to the clusters
at $z~\sim~0.55$.  Studies by Trentham (1998c) and Boyce et al. (2001) have 
shown that dwarf irregular (dIrr) galaxies have $B-R~\sim~0.9$, while dEs have colours in a broader 
range $1.3~\lsim~B-R~\lsim~2.0$. To compare the colours of the dwarf population in MS1008--1224 
with other nearby clusters, with more limited colour information,
we therefore plot in Fig. \ref{Figure7} the rest-frame $B-R$ colours of galaxies with 
$-17.7~<~M_B~<~-16.2$.

In the $B-R$ histogram of MS1008--1224 in Fig. 
\ref{Figure7} there are two peaks, suggesting that there indeed 
are two distinct populations representing dIrrs and dEs. In the figure we also show 
the distribution of $B-R$ colours for the faint galaxies, which on the basis of 
the full $BVRI$ photometry we have classified as early-type and late-type galaxies, respectively. 
The fact that this classification closely follows the two-colour division into the two 
peaks shows that we reliably can use the $B-R$ colour to broadly distinguish between early-type 
and late-type galaxies.

From the $B-R$ colours Trentham (1998c) finds that all 
dwarf galaxies with $-18.9~<~M_R~<~-16.9$ in Abell 665 at $z~=0.18$ are 
consistent with being dEs (called dSphs by Trentham). In Abell 963 at 
$z~=~0.21$ the majority of the dwarf galaxies with 
$-19.2~<~M_R~<~-18.2$ also have colours consistent with dEs. There 
is, however, in this cluster also a population with colours intermediate between dEs 
and dIrrs, which Trentham proposes may represent a transitional stage between these types.

In a recent study Boyce et al. (2001) investigate the $B-R$ colour of galaxies with 
$-19~<~M_R~<~-16.5$ in Abell 868 at $z~=0.154$. They find a distribution 
in colour with two peaks, showing the presence of a dominant population of dEs, but 
also a population of dIrrs. For the faintest galaxies ($-17.5~<~M_R~<~-16.5$) there 
are similar numbers of dEs and dIrrs.

The main difference between MS1008--1224 and lower-redshift clusters is the relative 
fraction of dIrrs to dEs. To make a quantitative comparison between the faint populations, 
we use the $B-R$ index to calculate the fraction of dIrrs for MS1008--1224, 
Abell 963, Abell 665 and Abell 868. We assume that galaxies with rest-frame $B-R~<~1.2$ 
are dIrrs, which is the criterion used by Boyce et al. (2001). 

A complication is that 
MS1008--1224 is not observed in standard Johnson--Cousins filters, but in Bessel filters. 
To correct for this we use galaxy templates from Coleman et al. (1980) and filter 
transmission curves to calculate the transformation between the standard $B-R$ colours 
and the $B-R$ colours for the Bessel filters. We then find that the division between dIrrs 
and dEs at $B-R$ = 1.2 in standard Johnson--Cousins filters corresponds to $B-R$ = 1.34 
in the Bessel filters. 

Using this criterion, we find that the fraction of faint galaxies 
that belong to the dIrr population is $0.72~\pm0.03$ for MS1008--1224, to be compared with 
$0.07~\pm0.03$ for Abell 963, $0.01~\pm0.01$ for Abell 665 and $0.22~\pm0.02$ for Abell 
868. A further problem when comparing these numbers is that the different investigations use 
different magnitude intervals for defining the dwarf populations. Irrespective of this, 
however, 
we reach the important conclusion that MS1008--1224 is dominated by dIrrs, while the 
lower redshift clusters are dominated by dEs.

The fraction of dIrrs in Coma, classified by the same colour criterion as above, for 
galaxies with $-15.3~<~M_R~<~-13.3$ is $0.11~\pm0.01$ 
(calculated from fig. 7 in Trentham 1998a). The fraction of {\it morphological} dIrrs in
Virgo ($-17.8~<~M_B~<~-15.8$), calculated from fig. 6 in Trentham 
\& Hodgkin (2002), is 0.25$\pm0.07$
The results emphasize the conclusion that dEs dominate the faint 
population in nearby clusters, while our study shows that the opposite is true
for MS1008--1224 at $z~=~0.31$. It would obviously be of great interest to study
more clusters to these limits to see if this is a general property.

Galaxy harassment could here be a mechanism that transforms the faint late-type population,
dominating medium redshift clusters such as MS1008--1224, into dEs, as shown by Conselice et al. (2001).
The similarity between the faint-end slopes of
the late-type population ($\alpha_f~=~-1.84\pm0.12$) and the early-type population 
($\alpha_f~=~-1.69\pm0.14$) at faint magnitudes ($M_B~>~-19$) is 
consistent with this picture. Based on these observations alone, we cannot, however, rule out
that slower mechanisms such as 'strangulation' (Balogh et al. 2000), are at least partly responsible
for the luminosity and colour evolution.

\subsection{Dwarf-to-giant ratio}
Previous studies have shown that the ratio of faint to bright 
galaxies increases at large radii, where the surface density decreases. Driver et al. (1998) 
find that five out of seven clusters at $z~\sim~0.15$ 
show an increasing dwarf-to-giant ratio with radius. A $K$--band study of AC 118 at 
$z~=~0.3$ by Andreon (2001) shows the same trend. In DFN02 we found that Cl1601+42 
has a steeper faint-end slope of the LF in the outer part of the cluster, which is 
equivalent to an increasing dwarf-to-giant ratio. 

Here, we calculate the dwarf-to-giant ratio by defining galaxies with 
$-19.5~<~M_B~<~-17.7$ as dwarfs, and brighter galaxies as giants. The separation
between dwarfs and giants is chosen to match the cross-over in the 
cluster LFs between early-type and late-type galaxies (Fig. \ref{Figure5}). This division
is somewhat brighter than used in other papers (e.g., Driver et al. 1998), 
and is here adopted in order to get sufficient statistics in the faint bin. We further divide 
each cluster into a core 
region with projected radius $<~0.5$ Mpc, and an outer region between 0.5 and 1.1--1.5 Mpc,
depending on cluster. 

\begin{figure}
\centerline{
\psfig{figure=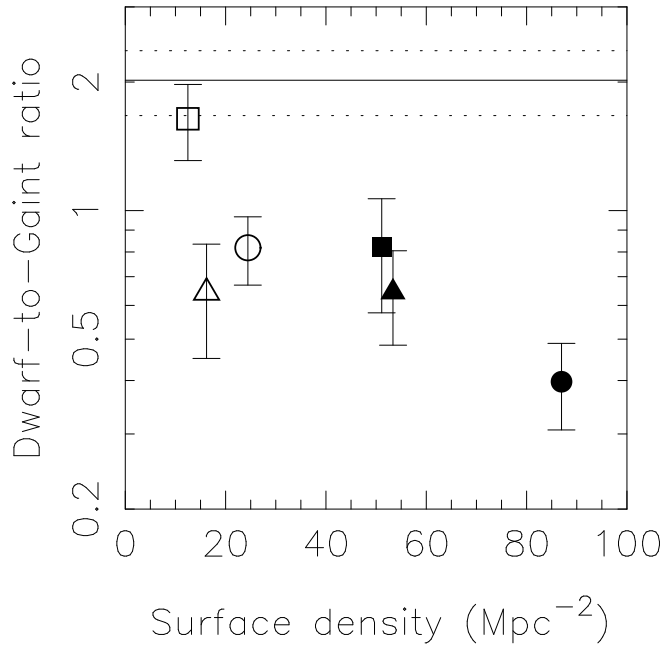,height=8.5cm}}
\caption{Dwarf-to-giant ratio as a function of bright galaxy surface density for Cl0016+16 
(circles), Cl1601+42 (squares) and MS1008--1224 (triangles). Filled symbols represent the 
inner region of the cluster ($R~<~0.5$ Mpc), while open symbols represent the outer region. 
The horizontal lines represent the field value (solid line) and 1$\sigma$-errors (dotted 
lines).}
\label{Figure8}
\end{figure}

In Fig. \ref{Figure8} we plot the resulting dwarf-to-giant 
ratios as function of surface density of the giant galaxies for these two regions in all 
clusters. Cl0016+16 is represented by circles, Cl1601+42 by squares and MS1008--1224 by 
triangles. Filled symbols show results for the core area, while open symbols represent 
results from the outer area. The horizontal lines represent the field value and 
the corresponding 1$\sigma$-errors.

Fig. \ref{Figure8} shows that Cl0016+16 and Cl1601+42 have a clear trend of an 
increasing dwarf-to-giant ratio when going from the inner high-density region, to the 
outer lower-density region, which is consistent with the trend found by Phillipps et 
al. (1998) and with Andreon (2001). As an explanation of this these authors suggest 
that the high-density environment in cores of clusters is 
hostile to the dwarf galaxies, while the outer, low-density regions do not affect 
this population, which therefore have a dwarf-to-giant ratio and faint-end 
slope similar to that of the field.
 
For the low redshift cluster MS1008--1224, there is no obvious trend. We note, however, 
that the error in the dwarf-to-giant ratio is substantial.
\subsection{The Butcher--Oemler effect as function of radius and limiting magnitude}
The BO effect is of major observational importance as a probe of the cluster galaxy population.
Although much of the discussion below to a large extent is a consequence of the early-type
and late-type LFs in Fig. \ref{Figure5} and their dependence on radius 
(e.g., Fig. \ref{Figure4}), we will for this reason illustrate the
consequences of this with two important implications for the BO effect.

\begin{figure}
\centerline{
\psfig{figure=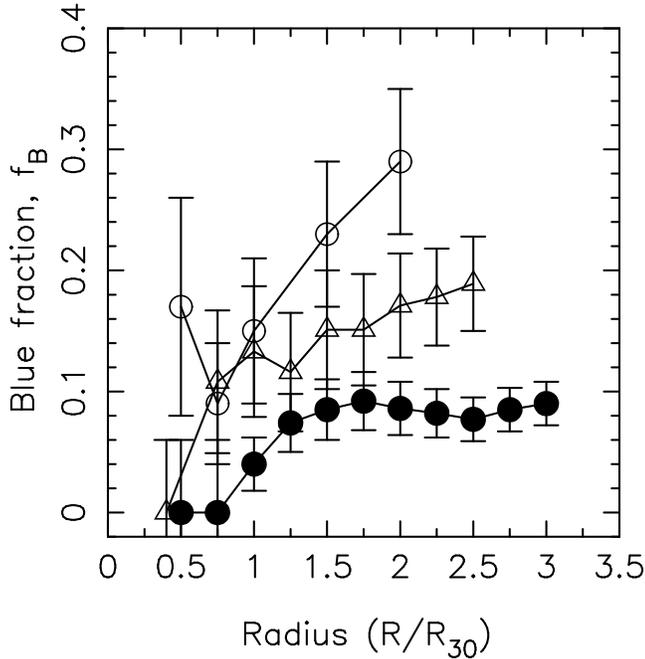,width=8.5cm}}
\caption{The blue fraction, $f_B$, as a function of radius for Cl0016+16 (filled circles), 
Cl1601+42 (open circles) and MS1008--1224 (triangles). $R_{30}$ is the fiducial value used 
by BO84 for calculating $f_B$.}
\label{Figure9}
\end{figure}

Already in BO84, and more recently in DFN02, it was pointed out that the blue fraction $f_B$ 
increases at larger radii. 
This can be understood as a result of the fact that early-types are in general more 
centrally concentrated than late-types. In Fig. \ref{Figure9} we show the dependence 
of $f_B$ on radius, where the radius is given in fractions of $R_{30}$ (as defined in
Section 4.3).
Also in this respect Cl0016+16 deviates from the other clusters.
Inside of 1.25$R_{30}$, $f_B$ decreases rapidly, 
while outside this radius $f_B$ stays almost constant, at a value 
well below that of the other clusters. Cl1601+42 on the other hand 
shows a steep radial dependence. 
The higher blue fraction in Cl1601+42 is consistent with the higher 
fraction of 
late type galaxies in this cluster (Fig. \ref{Figure4}). MS1008--1224 has a trend 
in between the two higher redshift clusters. From a very low $f_B$ in the core region, 
there is a rapid increase to $\sim~1.25~R_{30}$, followed by a slower increase in the 
outer regions. The behaviour of $f_B$ in these clusters reflects, as expected, the 
radial dependence 
of the early-type fraction shown in the right panel of Fig. \ref{Figure4}. 

\begin{figure}
\centerline{
\psfig{figure=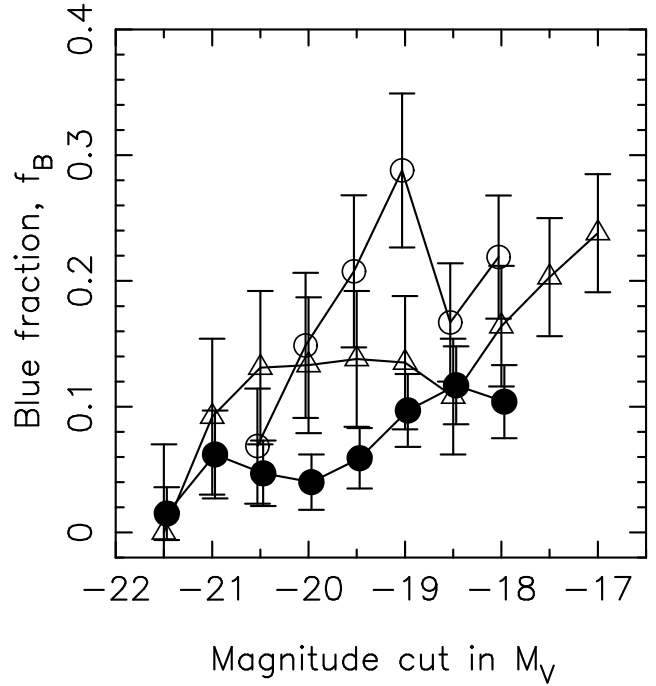,width=8.5cm}}
\caption{The fraction of blue galaxies, $f_B$, as a function of limiting magnitude for 
Cl0016+16 (filled circles), Cl1601+42 (open circles) and MS1008--1224 (triangles).}
\label{Figure10}
\end{figure}

In DFN02 we showed that $f_B$ in Cl1601+42 depends strongly of the limiting magnitude. 
This dependence is a direct consequence of the different shapes of the early-type and 
late-type LFs in this cluster. At fainter magnitudes the total LF becomes more dominated 
by late-type galaxies, which leads to a higher $f_B$. In Fig. \ref{Figure10} we plot the 
dependence of $f_B$ on the limiting magnitude for the three clusters. As expected, the 
increase of $f_B$ is much weaker in Cl0016+16 than in Cl1601+42, which is consistent 
with the LF for this cluster being dominated by early-type galaxies to fainter 
magnitudes (Fig. \ref{Figure5}). MS1008--1224 again shows an intermediate behaviour.
The strong increase in $f_B$ at faint magnitudes in MS1008--1224 is caused by the dominating
population of blue dwarf galaxies.

\section{Summary}

This work represents one of the deepest studies of the LF for
intermediate-redshift clusters, several magnitudes beyond $L^*$.  In
addition, we have demonstrated the usefulness of photometric redshifts
for selecting clusters members, as well as for population studies of
distant clusters of galaxies. As a result, while most previous
determinations of distant cluster LFs have treated only the {\it
total} LF, we have in this study been able to separate the early-type and
late-type populations. This has been possible, despite the moderate
telescope size due exactly to the use of photometric
redshifts. Compared to the usual subtraction method, we have been
gaining a large factor in respect to the background contamination.

Our main conclusions from this study are:

\begin{itemize}

\item
There is no universal shape of the {\it total} cluster LF at $z~\gsim~0.3$.

\item
The early-type population has a Gaussian LF, while the late-type population is
well fitted by a Schechter function. This suggests that the LFs for different 
spectral populations could be universal, while the total LF depends on the relative abundance 
of these populations.

\item
The evolution of the late-type galaxies is consistent with a fading by $\sim$ 2 
magnitudes between $z~\sim~0.55$ and $z~=~0$, while the early-type population fades
by $\sim~1$ mag. This scenario suggests that the total LFs of the high-$z$ clusters 
become more similar to local LFs as the clusters get dynamically older. 

\item The red cluster Cl0016+16 is an atypical high-$z$ cluster that resembles local 
rich clusters in many aspects, indicating an old dynamical age despite its redshift.
It does, however, contain a large fraction of post-starburst galaxies, suggesting that
star formation was more intense at $z~\gsim~0.8$, and that the infall was very low during the last 
$\sim~1$ Gyr before the cluster is observed. 

\item
In MS1008--1224 at $z~=~0.31$, we find that dIrrs dominate over dEs, opposite to what is found in nearby clusters. If this is confirmed for more clusters at this redshift it implies a dramatic evolution of the dwarf population.

\item
The relation between dwarf-to-giant ratio and surface density indicates that 
high-density regions are hostile to dwarfs, consistent with a destruction or fading of 
this population by galaxy harassment.

\item We find that the blue fraction, $f_B$, as defined by BO84
varies with radius and limiting magnitude. This is a direct consequence of the 
radial gradient of the late-type galaxies and the relative normalisation of the 
late-type and early-type LFs.

\end{itemize}

There are several natural next steps to this study. The sample needs
to be expanded, both in terms of redshift and cluster properties. As
the example of Cl 0016+16 has shown, even at the same redshift the the
population may vary greatly, depending on the dynamical state, and
possibly on the environment. A better understanding of this is of
obvious importance for the  understanding of the cluster-formation
process, especially at redshifts higher than we have
probed. 

Spectroscopic studies at redshifts similar to the clusters in this
study may today marginally be carried out with 8--10 m class
telescopes. This would of course give more detailed information about
the dynamical properties of the clusters, as well as a general check
of the reliability of the photometric redshift method for clusters. At
higher redshifts it is, however, virtually impossible to probe the
population beyond L$_*$ with spectroscopy, even with these
telescopes. For this type of investigations one therefore has to rely on
photometric redshifts, in this case extended to include the near-infrared
bands. The study in this paper, as well as DFN02, hopefully represents
some first steps in this direction.

Also at low redshifts there are several open questions, especially
connected to the dwarf population. The dramatic difference in the dIrr-to-dSph 
ratio seen between MS1008--1224 and the local clusters needs to be
confirmed with more medium-redshift clusters.

Finally, our study has mainly given information about the broad band
colours of the cluster members. A high-resolution study with HST of
the morphological properties of these galaxies would be of prime
interest, as was partially demonstrated for Cl 1601+42 in DFN02,
and in more detail for the bright population by e.g. the MORPHS sample
(Smail et al. 1997). We hope in the future to pursue some of these
questions.

\section*{Acknowledgments}
We are grateful to Neil Trentham and Claes-Ingvar Bj\"{o}rnsson for several useful comments
and to the referee for many valuable comments and suggestions.
Nordic Optical Telescope is operated on the island of La Palma 
jointly by Denmark, Finland, Iceland, Norway, and Sweden, in the
Spanish Observatorio del Roque de los Muchachos of the 
Instituto de Astrofisica de Canarias.

The data presented here have been taken using ALFOSC, which is owned by the Instituto de 
Astrofisica de Andalucia (IAA) and operated at the Nordic Optical Telescope under agreement 
between IAA and the NBIfAFG of the Astronomical Observatory of Copenhagen.

This paper is also based on observations obtained at the Very Large Telescope
at Cerro Paranal operated by the European Southern Observatory.
 
This work has been supported by the Swedish Research Council.

\label{lastpage}
\end{document}